\documentclass{emulateapj}
\shorttitle{Spectroscopy of Bo\"otes~II} 
\shortauthors{Koch et al.} 
\slugcomment{Accepted for publication in the Astrophysical Journal}

\begin{document}

\title{A spectroscopic confirmation of the Bo\"otes~II dwarf spheroidal\altaffilmark{1}}

\author{Andreas Koch\altaffilmark{2}, Mark I. Wilkinson\altaffilmark{3}, Jan T. Kleyna\altaffilmark{4},  
Mike Irwin\altaffilmark{5}, Daniel B. Zucker\altaffilmark{5}, \\ Vasily Belokurov\altaffilmark{5}, Gerard F. Gilmore\altaffilmark{5}, 
 Michael Fellhauer\altaffilmark{5,6}, and N. Wyn Evans\altaffilmark{5}}
\email{akoch@astro.ucla.edu}
\altaffiltext{1}{Based on observations made with the Isaac Newton Telescope operated on the island of 
La Palma by the Isaac Newton Group in the Spanish Observatorio del Roque de los Muchachos of the 
Instituto de Astrof'sica de Canarias.}
\altaffiltext{2}{UCLA, Department of Physics \& Astronomy, 430 Portola Plaza, Los Angeles, CA 90095, USA}
\altaffiltext{3}{Dept. of Physics and Astronomy, University of Leicester, University Road, Leicester LE1 7RH, UK}
\altaffiltext{4}{Institute for Astronomy, Honululu,  2680 Woodlawn Drive, Honolulu, HI  96822, USA}
\altaffiltext{5}{Institute of Astronomy, Madingley Road, Cambridge, CB3 0HA, UK}
\altaffiltext{6}{Departamento de Astronomia, Universidad de Concepcion, Casila 160-C,
Concepcion, Chile}

\begin{abstract}
We present a new suite of photometric and spectroscopic data for the
faint Bo\"otes~II dwarf spheroidal galaxy candidate. Our deep
photometry, obtained with the INT/WFC, suggests a distance of 46 kpc and a 
small half-light radius 
of 4.0$\arcmin$ (56 pc),
consistent with previous estimates. Follow-up spectroscopy obtained
with the Gemini/GMOS instrument yielded radial velocities and
metallicities.  While the majority of our targets covers a broad range
in velocities and metallicities, we find five stars which share very
similar velocities and metallicities and which are all compatible with
the colors and magnitudes of the galaxy's likely red giant branch. We
interpret these as a spectroscopic detection of the Bo\"otes~II
system. These stars have a mean velocity of $-$117 km\,s$^{-1}$, a
velocity dispersion of (10.5$\pm$7.4) km\,s$^{-1}$ and a mean [Fe/H]
of $-1.79$ dex, with a dispersion of 0.14 dex.  At this metallicity,
Boo~II is not consistent with the stellar-mass-metallicity relation
for the more luminous dwarf galaxies.  Coupled with our distance
estimate, its high negative systemic velocity rules out any physical
connection with its projected neighbor, the Bo\"otes~I dwarf
spheroidal, which has a velocity of $\sim +100$ km\,s$^{-1}$.  The
velocity and distance of Bo\"otes~II coincide with those of the leading
arm of Sagittarius, which passes through this region of the sky, so
that it is possible that Bo\"otes~II may be a stellar system
associated with the Sagittarius stream. Finally, we note that the
properties of Bo\"otes~II are consistent with  being the surviving
remnant of a previously larger and more luminous dSph
galaxy. \end{abstract}
\section{Introduction}
Dwarf spheroidal galaxies (dSphs) are a well-established mainstay in
discussions of cosmological structure formation.  Although these
small-scale systems are clearly dark matter dominated (Mateo 1998;
Gilmore et al.  2007), the properties of the more luminous dSphs are
difficult to reconcile with a simplistic building block scenario
(e.g., Unavane et al. 1996; Moore et al. 1999; Venn et al. 2004).  In
this context, proposed solutions to major controversies such as the missing
satellite problem (Robertson et al. 2005; Font et al. 2006; Strigari
et al. 2007; Simon \& Geha 2007; Bovill \& Ricotti 2008) have been
further fueled by a plethora of discoveries of even fainter dSph
candidates around the Milky Way using data from the Sloan Digital Sky
Survey (SDSS; Willman et al. 2005a,b; Zucker et al. 2006; Belokurov et
al. 2006a, 2007; Walsh et al. 2007a). Characteristic properties of
these systems are very low luminosities and low stellar masses, while
the apparently high mass-to-light ratios (of up to several hundreds) 
seen in several of them (Simon \& Geha 2007) are usually taken as indicative of a
dominant dark matter component.

Gilmore et al. (2007) showed that there is a minimum half-light radius
of $\sim100$pc for all isolated dSphs more than 50kpc from the
Galactic centre, and a maximum half-light radius of equilibrium (dark matter-free)
globular clusters (GCs) of $\sim30$pc. They suggested this minimum size is
an intrinsic feature of systems with dark matter halos. For this to be
true, the few very-low luminosity objects in or close-to the
corresponding size gap would be tidally disrupting/disrupted dSphs or
star clusters. The radius distribution has been confirmed in a
subsequent analysis of SDSS photometric data (Martin et al. 2008). The
characterisation of the few observed systems of sizes intermediate
between dSphs and star clusters is thus a test of the minimum size relation 
proposed by Gilmore et
al. (2007). To show that such a system is a
former dSph would require detecting a residual or truncated dark
matter halo and/or a large intrinsic chemical abundance dispersion.

Among the new ultra-faint objects is Bo\"otes~II (hereafter Boo~II),
which was identified as a stellar overdensity on the sky by
Walsh et al. (2007a) and which exhibits a vague main sequence turnoff
(MSTO) and only a sparse red giant branch (RGB).  Boo~II is among the
four faintest dSph candidates presently known ($M_V=-2.7$; Martin
et al. 2008).  Moreover, its spatial extent ($\sim$36--50 pc; Walsh et al. 2007b; 
Martin et al. 2008) renders it incompatible with a
classical GC which has been unaffected by tides.
Its small projected separation (only $1.6\degr$)  from the faint
Bo\"otes~I dSph and its radial distance of 42 kpc (Walsh et al.  2007b),
close to that of Bo\"otes~I (ca. 60 kpc; Belokurov et al. 2006a; Siegel
2006; Dall'Ora et al. 2006) prompted the suggestion that these two
systems might be in some way associated. However, despite a mild
degree of distortion in the isophotes of Bo\"otes~I, there is no current
evidence of any interaction between these systems, which might be
expected to give rise to aligned elongations in each dSph (see also
Fellhauer et al. 2008).  Although photometric studies suggest that
Boo~II is an old, metal-poor system (Walsh et al. 2007a,b), nothing is
yet known about its kinematics.

In this work we investigate the physical nature of Boo~II based on
deep photometry and spectroscopic follow-up of this stellar system, by
means of which we aim to measure its velocity and metallicity.  In
this way we can determine whether there is any possible connection
between the pair of Bo\"otes and constrain some of
the important properties of Boo~II.
This Paper is organized as follows: in \textsection2, the photometric
and spectroscopic data and their reduction are described, while
\textsection3 focuses on the derivation of radial velocities and 
metallicities, based on which we claim a spectroscopic detection of
Boo~II. After a discussion of the possible nature of Boo~II in
\textsection4 we summarize our findings in \textsection5.
\section{Data and Reduction}
\subsection{Photometry}
Boo~II was observed in sub-arcsec seeing conditions on the night of
2007 May 11 using the Wide Field Camera on the 2.5-m Isaac Newton
Telescope (INT) on La Palma. The images were corrected for
non-linearity, bias-corrected, trimmed, flat-fielded and defringed (i'
only) using the processing pipeline developed by Irwin \& Lewis
(2001).  This software was then used to generate detected object
catalogues for each individual exposure (3 $\times$ 900s for g' and 3
$\times$ 600s for i'). The catalogues were used to refine the World
Coordinate System information for each image prior to forming the deep
image stacks and subsequently-generated deep catalogues that form the
basis for the rest of this analysis.  A preliminary photometric
calibration was based on Landolt (1992) standard fields observed
throughout the night.  Since the night was partially non-photometric,
due to varying levels of dust extinction, we then bootstrapped the
data onto the SDSS photometric system using a table of SDSS stars from
that region. In the SDSS AB system, the stacked data reach a 5$\sigma$
depth of approximately 25.5 in g and 24.0 in i.

Figure~1 shows the resulting color-magnitude diagrams (CMDs) for the central 
five arcmin of Boo~II (corresponding to about 1.2 half-light radii $r_h$; e.g., 
Martin et al. 2008) and an offset control field  located 
15.6$\arcmin$ (3.7 $r_h$) from the center of Boo~II on an adjacent CCD (\#2) 
detector. The right-hand panel of the figure shows the luminosity function 
(LF) in the central field together with a comparison LF derived from a larger 
area of CCD \#2 and normalised to the same area as the
central region.  While there is a clear excess of stars in the main sequence
and turn-off regions ($g-i \sim0.25$, $21.5<i<24.0$), at brighter magnitudes  
where the sub-giant and RGB locii are expected, and from where we draw our
spectroscopic targets, there is a much weaker excess.

\begin{figure}
\begin{center}
\includegraphics[width=1\hsize]{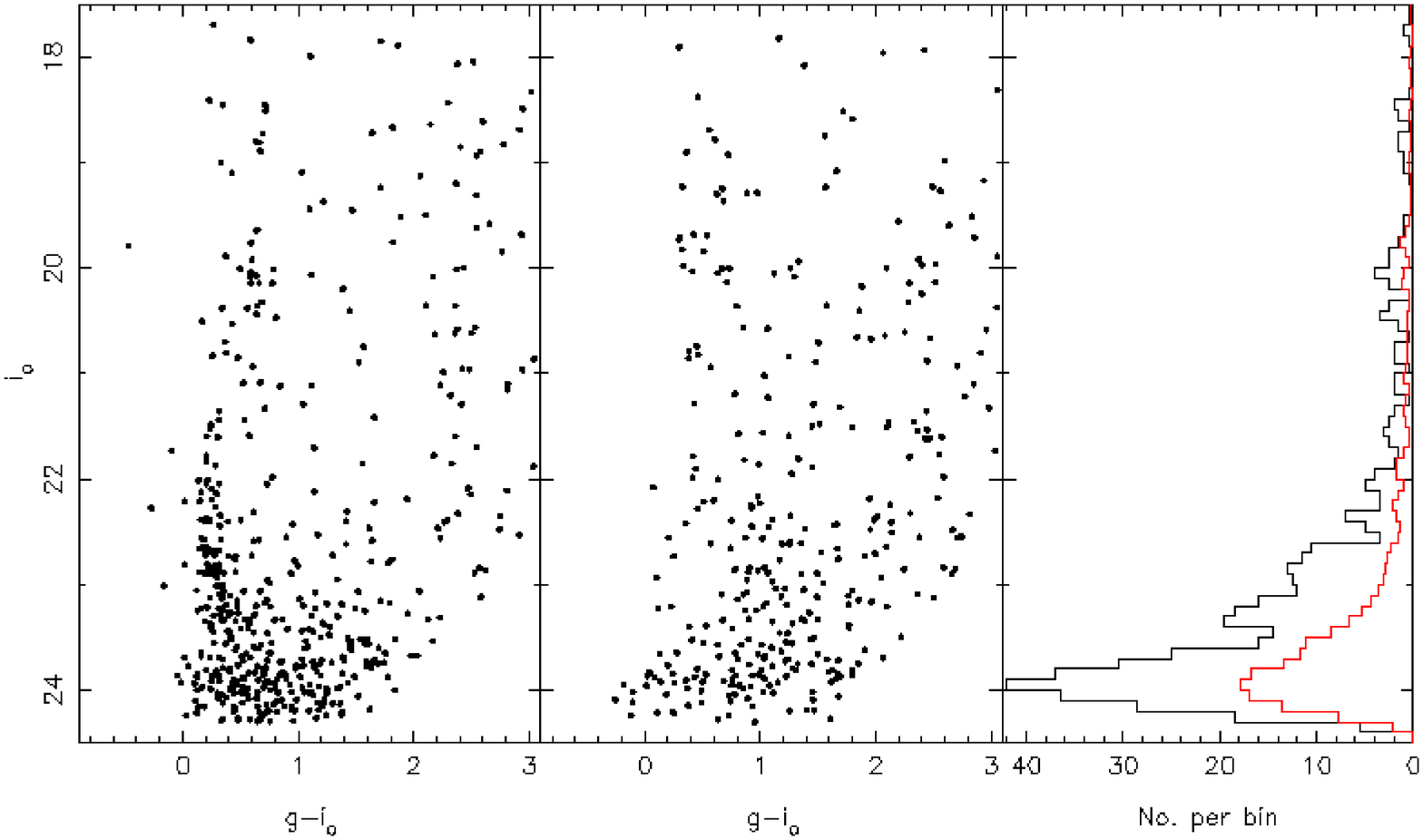}
\end{center}
 \caption{{\em Left panel:} ($g-i$, $i$) CMD  for stars
 within five arcmin of the center of Boo~II. {\em Middle:} CMD 
 for stars in a field offset by 15.6$\arcmin$ from the
 centre of Boo~II covering the same area as that of the central
 field. The right panel shows  the $i$-band luminosity functions of
 stars in the central field (black) and from a much larger offset region 
 (red line) suitably area-normalised.  There is a clear excess of stars on 
 the main sequence of Boo~II and a smaller but still significant excess for 
 turn-off and sub-giant stars.}
\end{figure}

\subsection{Spectroscopy}
We obtained spectra for seventeen stars in the field of Boo~II using a single
pointing of the GMOS-N spectrograph, mounted on the Gemini North
telescope, on 2007 April 17. Our targets were chosen by cross-matching
of GMOS-N $i$-band pre-images with existing SDSS photometry. All target
stars were selected to lie in the region of the CMD occupied by the sparse RGB of Boo~II 
(see Figures~1,6). 
These are listed in Table~1 with their characteristic 
properties.
\begin{deluxetable*}{rccccrrccrcc}
\tabletypesize{\scriptsize}
\tablecaption{Characteristics of target stars. Magnitudes are those from the SDSS. 
}
\tablewidth{0pt}
\tablehead{
\colhead{}  &   \colhead{$\alpha$} & \colhead{$\delta$} &  \colhead{}& \colhead{} & 
\colhead{v$_{\rm HC}$} & \colhead{$\sigma$\,v} & \colhead{} & \colhead{} & \\
 \raisebox{1.5ex}[-1.5ex]{ID} & \multicolumn{2}{c}{(J2000.0)} & \raisebox{1.5ex}[-1.5ex]{i} & 
 \raisebox{1.5ex}[-1.5ex]{g$-$i} &\multicolumn{2}{c}{[km\,s$^{-1}$]}  & \raisebox{1.5ex}[-1.5ex]{[Fe/H]} &
 \raisebox{1.5ex}[-1.5ex]{$ \sigma$\,[Fe/H]} &\raisebox{1.5ex}[-1.5ex]{S/N} 
 & \raisebox{1.5ex}[-1.5ex]{Quality} & \raisebox{1.5ex}[-1.5ex]{Membership}
}
\startdata
 1 & 13:58:07.2 & 12:51:47.5 & 19.04 & 0.41 & $-$165.55 &  4.73 & $-$1.53 &  0.12 & 17 &  \\
 2 & 13:58:08.6 & 12:51:15.4 & 20.05 & 0.55 & $-$131.49 &  8.48 & $-$1.72 &  0.14 & 12&& Boo~II  \\
 3 & 13:58:09.5 & 12:51:25.5 & 19.83 & 0.67 & $-$109.38 &  5.28 & $-$1.74 &  0.12 & 14 & & Boo~II \\
 4 & 13:58:08.1 & 12:53:53.1 & 19.99 & 0.96 &	  35.49 &  5.39 & $-$0.84 &  0.18 & 10 & \\
 5 & 13:58:10.5 & 12:55:41.8 & 20.25 & 0.85 &	  66.00 &  15.1 & $-$0.80 &  0.22 & 8 & \\
 6 & 13:58:05.9 & 12:51:12.2 & 22.38 & 0.14 &  $-$88.86 &  15.0 &   \dots & \dots & 3 & s \\
 7 & 13:58:01.5 & 12:51:04.7 & 18.83 & 0.75 & $-$125.58 &  4.07 & $-$1.99 &  0.10 & 21 && Boo~II  \\
 8 & 13:58:00.7 & 12:53:45.0 & 19.14 & 0.52 &	  95.93 &  3.17 & $-$1.24 &  0.13 & 17 & \\
 9 & 13:57:59.8 & 12:54:26.1 & 18.52 & 0.84 & $-$123.08 &  1.87 & $-$1.61 &  0.11 & 28 & & Boo~II \\
10 & 13:58:00.3 & 12:55:39.9 & 17.51 & 1.05 &	$-$4.44 &  1.76 & $-$0.82 &  0.13 & 55 & \\
11 & 13:58:02.6 & 12:51:35.9 & 21.42 & 0.49 &  $-$90.57 &  28.8 &   \dots & \dots & 4 & n \\
12 & 13:58:01.9 & 12:53:45.9 & 22.00 & 0.82 &	 238.11 &  6.25 &   \dots & \dots & 3 & s \\
13 & 13:57:57.2 & 12:53:15.8 & 20.54 & 0.57 &	   4.26 &  7.70 & $-$1.45 &  0.39 & 7 & \\
14 & 13:57:58.4 & 12:53:31.5 & 21.42 & 0.70 & $-$125.05 &  9.86 & $-$1.02 &  0.23 & 4 & a \\
15 & 13:57:51.2 & 12:51:36.6 & 18.50 & 0.76 & $-$100.07 &  2.33 & $-$1.81 &  0.10 & 27 & & Boo~II \\
16 & 13:57:53.4 & 12:55:15.3 & 18.88 & 0.87 &  $-$10.50 &  2.24 & $-$0.57 &  0.15 & 20 & \\
17 & 13:57:52.2 & 12:52:47.1 & 22.13 & 0.88 &	 301.14 &  9.13 &   \dots & \dots & 3 & n 
\enddata
\tablecomments{Quality flags are: (a) Ambiguous CCF peak;  (n) No CCF peak discernible;  
(s) Spurious velocity due to sky residuals. Boo~II member candidates are marked in the last column. 
See text for details.}
\end{deluxetable*}

A single GMOS slit mask was prepared with slitlets of width
0.75$\arcsec$. The spectra were centered on the spectral region
containing the \ion{Ca}{2} triplet (CaT) lines. Exposures were taken
at central wavelengths of both 8550 and 8600\AA~in order to achieve
continuous wavelength coverage in the spectra across the gaps between
the CCDs. Our observations used the R831+\_G5302 grating and
CaT\_G0309 filter, with 2$\times$2 binning in the spectral and spatial
dimensions. The spectra thus obtained have a nominal resolution of
3600. The total integration time was 12600\,s, which we divided into
individual exposures of 1800\,s to facilitate cosmic ray removal.

The raw data were reduced using the standard {\em gemini} reduction
package within the Image Reduction and Analysis Facility ({\sc
iraf}. CuAr
lamp exposures adjacent in time to the science exposures provided the
calibration frames. The typical r.m.s.  uncertainty in the wavelength
calibration, obtained by fitting a polynomial to the line positions in
the CuAr spectra, was 0.04\AA, which corresponds to a velocity
uncertainty of $\sim$1.4 km\,s$^{-1}$ at the CaT. Typically, the final
spectra have per-pixel signal-to-noise (S/N) ratios which range from
very low quality (S/N as low as 3--5) up to $\sim$25, with a median
S/N of 12.
Three sample spectra of stars covering a representative magnitude
range are shown in Fig.~2.
\begin{figure}
\includegraphics[width=1\hsize]{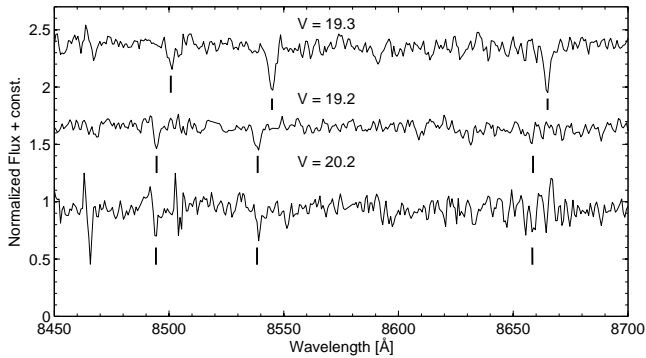}
 \caption{Sample spectra of a brighter foreground dwarf (top; S/N$\sim$17), 
 and a typical (middle; SN$\sim$21) and fainter 
 (bottom; SN$\sim$12) Bo\"otes~II giant candidate. 
The three CaT lines are indicated with vertical ticks. The Doppler shifts clearly  separate the 
foreground star from the Boo~II candidates.}
\end{figure}

\section{Analysis}
\subsection{Structural parameters}
Although two analyses of the structure of Boo~II have been published already
by Walsh et al. (2008) and Martin, de Jong \& Rix (2008), the uncertainty
in even the basic structural parameters is large enough to warrant further
examination.  Our relatively deep wide area photometry reaches well below 
the MSTO (see figure 1) and with a total coverage of 0.25 square degrees  
extends far enough from the main body of Boo~II to enable a reliable 
background estimate.  In particular, the published values for the 
half-light radius, $r_h$, central surface brightness, $\mu_0$, and absolute 
magnitude, M$_v$, emphasise the apparent hybrid nature of the object by 
locating it in the void between dwarf galaxies and stellar clusters
(e.g. Belokurov et al. 2007).

To help further characterize the structure of Boo~II we first constructed
an isopleth (number density) map by counting stars satisfying $21.5 < g < 24.5$
and $g-i < 0.6$ on a 15 arcsec grid over a 0.5 $\times$ 0.5 degree region.
We show in Fig.~3 a contour map of these number counts, smoothed with a 
Gaussian kernel of FWHM 1 arcmin.  The ``background'' level, 0.60 $\pm$0.05 
arcmin$^{-2}$, used here and for the subsequent profile analysis, was estimated
from the average counts on CCD \#2, the closest part of which is 11 arcmin to
the West of the centre of Boo~II.

\begin{figure}
\begin{center}
\includegraphics[width=1\hsize]{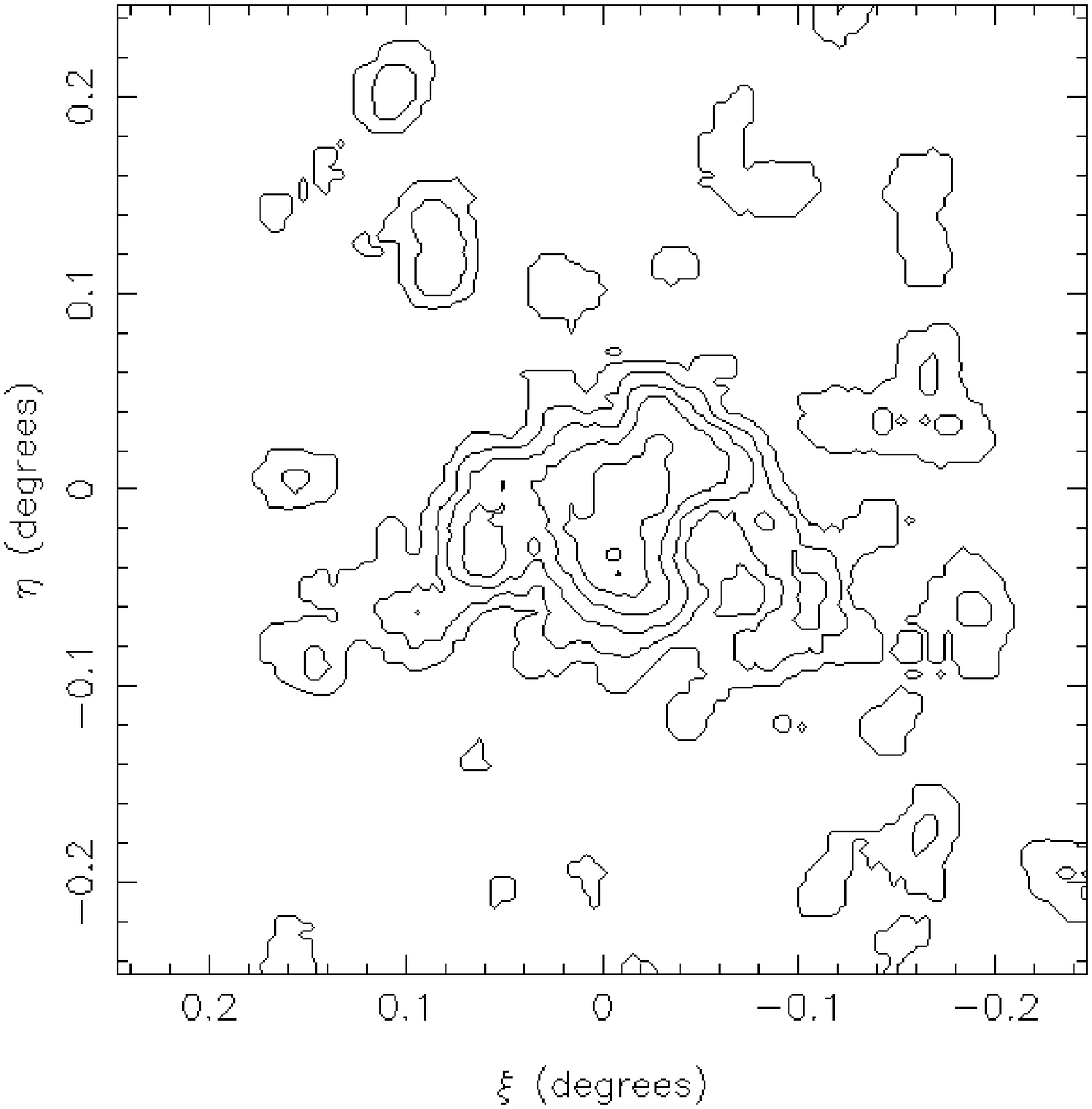}
\end{center}
 \caption{Contours of the number density of stellar objects satisfying 
 $21.5 < g < 24.5$ and $g-i < 0.6$, N to the top and E to the left.  
 These have been smoothed with a Gaussian of FWHM 1 arcmin and highlight 
 the irregular appearance of Boo~II. Contour levels begin at 0.5 arcmin$^{-2}$ 
 above background, in steps of 0.5 arcmin$^{-2}$.}
\end{figure}
The highly irregular shape of Boo~II renders an ellipticity and position angle 
estimate from our data difficult.  Martin et al. (2008), from a maximum
likelihood analysis, obtained values for ellipticity and position angle of 
0.21$\pm$0.21 and $-35\degr^{+48}_{-55}$, respectively, emphasising the 
uncertainty of imposing an elliptical morphology in a case like this.

We therefore decided to minimise the number of degrees of freedom and analyse 
the photometric properties of Boo~II using circular annuli centred on our
derived position of ($\alpha$,$\delta$)=(13$^h$58$^m$05$^s$, 12$^\circ$52'0'').  In this case
the background-subtracted radial profile, from annuli spaced by 1 arcmin,
is quite well-defined and is shown in figure~4 together with an overlaid model 
fit.  A Plummer law with a central number density of 3.5 arcmin$^{-2}$ and 
half-light radius $r_h = 4.0\arcmin^{+0.7}_{-0.3}$ adequately characterises the 
profile. This value is in good agreement with the $4.2\arcmin^{+1.1}_{-1.4}$ 
found by Martin et al. (2008), and somewhat larger than $r_h$ of 
2.5$\arcmin\pm 0.5\arcmin$ from Walsh et al. (2007b). 
The profile in figure~4 is clearly not well-described by either a power law
or an exponential, and, although a King model could no doubt be tailored
to give an adequate fit, the extra degree of freedom and/or concept of a tidal
radius for this system is hardly warranted.

\begin{figure}
\begin{center}
\includegraphics[width=1\hsize]{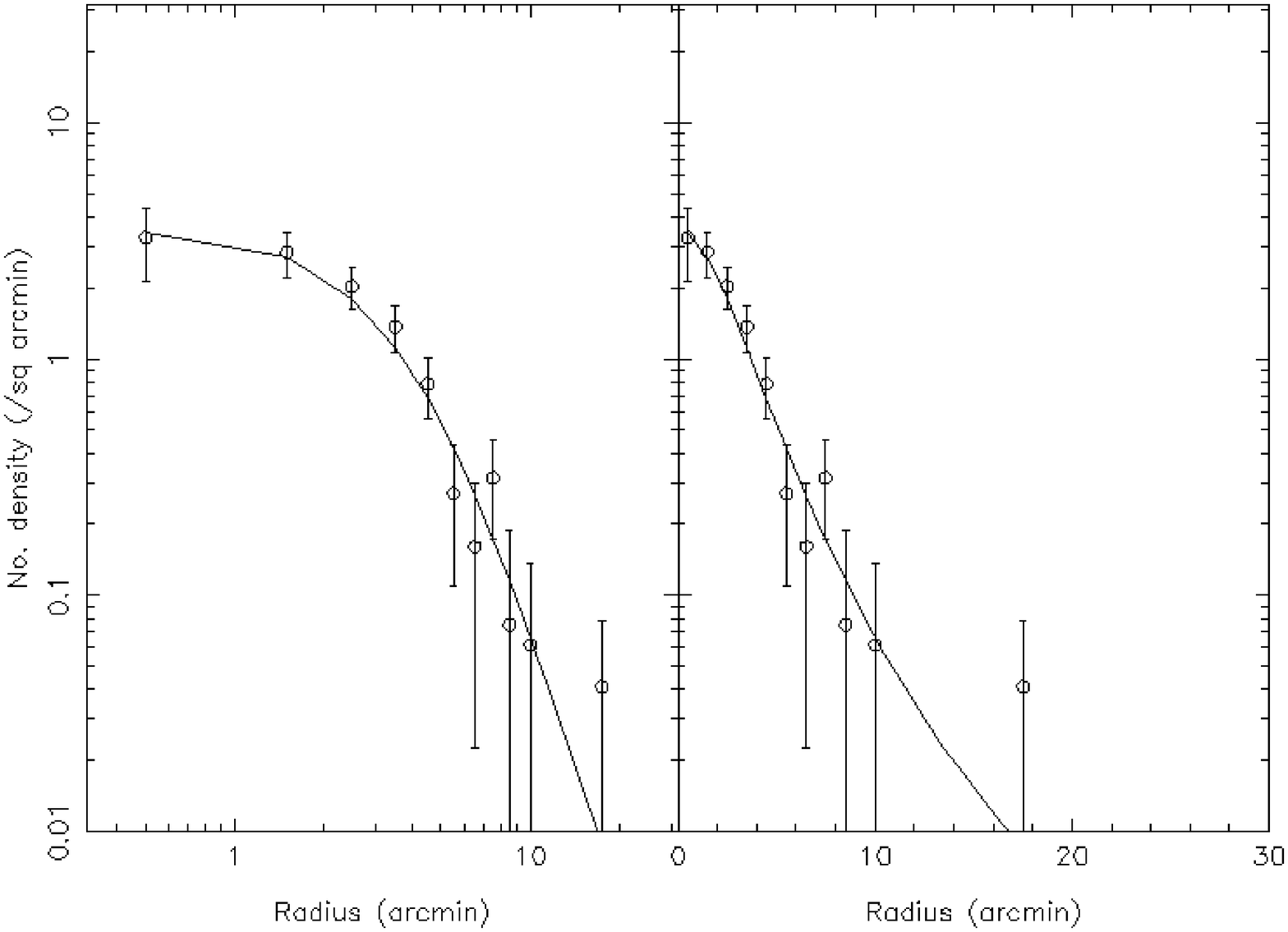}
\end{center}
 \caption{Background-corrected radial density profile of Boo~II with the 
 best-fit Plummer profile overlaid.  The error bars include contributions 
 from Poisson counting statistics and the error in the background estimate.}
\end{figure}

The sparsity of potential Boo~II CMD features brighter than the MSTO, its 
structural irregularity and the faintness of the system, make it difficult 
to derive an accurate direct estimate of the total magnitude.  In contrast, 
however, the central surface brightness of the CMD region used in the profile 
analysis is quite well-defined.  We therefore estimated what fraction of 
the total luminosity, and hence surface brightness, is contained in this 
region by comparison with the overall LF of M92 (Walsh et al. 2007a,b), 
assuming a similar stellar population for Boo~II.  Thus, by using only the 
well populated parts of the CMD (see Fig.~1) and re-scaling to correct for
missing subgiants, RGB and fainter main sequence stars, we derive a central
surface brightness in the Vega system of $\mu_{0,i} = 28.6$ and 
$\mu_{0,g} = 29.2$ mag\,arcsec$^{-2}$ with approximate error of $\pm$0.5
mag\,arcsec$^{-2}$.  Integrating the Plummer profile then gives estimates
of the total magnitude of  $M_i\sim-2.7$ and $M_g\sim-2.1$.  These are in 
good agreement with the values of $M_v = -2.7$ and $\mu_{0,V} = 28.5$ 
mag\,arcsec$^{-2}$ derived by Martin et al. (2008) and confirm the unusual
position of Boo~II in the $r_h$-versus-$M_v$ ($\mu_{0,V}$) domain.

\subsection{Radial velocities}
Radial velocities of our targets were determined by cross correlation
against a synthetic template, composed of the three CaT lines, using
{\sc iraf}'s {\em fxcor} task.  For the faintest stars, this procedure
did not yield any clear correlation peak. We therefore followed Zucker
(2003) in determining instead the cross correlation function (CCF)
from each of the seven individual exposures of each star and
subsequently combining the separate CCFs into a straight average CCF
for each target. This process efficiently increases our ability to
detect the underlying velocity signal. In practice, the relative
radial velocity was then determined from a Gaussian fit to the
strongest average-CCF peak.  The median radial velocity error on our
measurements, as determined from the covariance matrix of the CCF fit
(Zucker 2003), is 5.4 km\,s$^{-1}$.

The third CaT line at 8662\AA\ is prone to strong contamination by sky
line residuals.  While cross correlation against the third line only
did not result in any measurable peaks in the CCF for most of the
targets, its inclusion in the entire cross correlation region did not
affect the derived velocities -- the correlation of the entire CaT
from 8475--8680 \AA\ yields essentially the same velocities as a
correlation restricted to only the first two lines at
$\lambda\lambda$8498,\,8542 \AA.
For two of the faintest stars, no CCF was discernible at all and we
discard these from further analysis (see the quality flag in Table~1).
The low S/N spectra of the faint stars are particularly sensitive to
sky removal and will inevitably contain stronger sky residuals
compared to the weak absorption features.  A cross correlation of a
sky emission spectrum against the synthetic CaT template reveals that
the strongest sky lines will produce spurious CCF peaks at $-86$ and
+240 km\,s$^{-1}$, respectively (see also Kleyna et al. 2004).  It is
thus likely that the strong peaks at exactly these velocities in the
CCFs of the two targets with i-band magnitudes of 22.0 and 22.4 mag (stars \#6 and \#12) are
such sky residuals. We excluded these measurements from our sample as
well.
Finally, for one star the cross correlation yielded two peaks of
comparable height, at $\sim$$-120$ and +100 km\,s$^{-1}$. We flagged
this measurement as ``dubious'' (see Table~1) and continue by adopting
the negative velocity, but account for its uncertain nature by both
including and excluding it in the subsequent quantitative analyses.
Our final set of heliocentric radial velocities, v$_{\rm HC}$, is
listed in Table~1 and shown in the histogram in Figure~5.

\begin{figure}
\includegraphics[width=1\hsize]{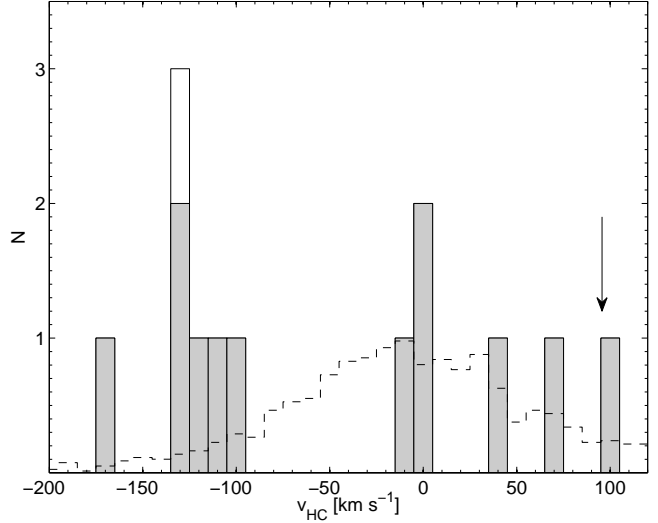}
 \caption{Velocity histogram of the 12 stars for which we could
 measure reliable velocities. The white bar indicates the target with
 only an ambiguous velocity signal in the CCF.  The expected
 foreground contamination from the Galactic Besan\c con model (Robin
 et al. 2003) is shown as a dashed histogram.  An arrow illustrates
 the systemic velocity of the Bo\"otes~I dSph.}
\end{figure}

Those stars with velocities above $-$20 km\,s$^{-1}$ are certainly
Galactic foreground stars as a comparison with the Galactic Besan\c
con model (Robin et al. 2003) confirmed (dashed line in Figure~5).
While there is still a non-negligible fraction of Galactic
contaminants expected below $\sim$$-100$ km\,s$^{-1}$, we note the
presence of an excess of stars at $\sim$$-120$km\,s$^{-1}$, which
suggests a probable kinematic detection of Boo~II member stars. If
these were Galactic halo stars (as predicted by the 
model), it is curious that we do not see a more symmetric velocity 
distribution in our data; this again argues in favor of our having detected Boo~II as
an independent system. Moreover, a Kolmogorov-Smirnov test showed that the 
hypothesis that all stars with good velocities are from the same distribution as the Galactic model stars 
can be rejected at a 98\% confidence level (2.3$\sigma$), while the 
sample {\em without} the Boo~II candidates at $-$120 km\,s$^{-1}$ is compatible with the 
foreground distribution at 35\%, which again suggests that the stars in the negative velocity peak 
are likely not Galactic foreground. 

\subsection{Isochrone fits}
In Figure~6 we show the location of our target stars in
color-magnitude space.  Stars belonging to the low-velocity peak are
flagged as potential members (solid circles), while the likely
foreground contaminants are shown as open symbols.
\begin{figure}
\begin{center}
\includegraphics[width=1\hsize]{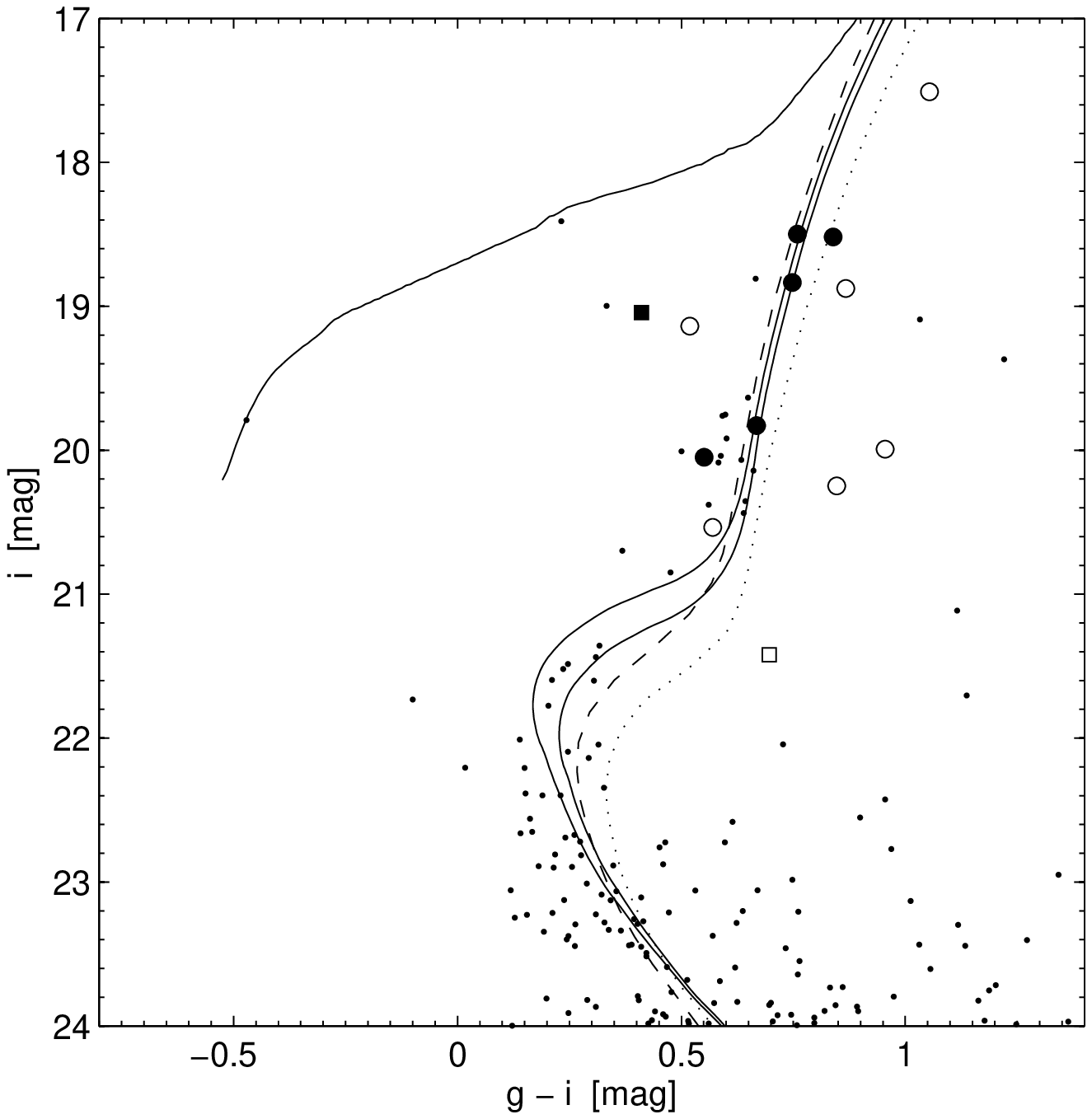}\\
\includegraphics[width=1\hsize]{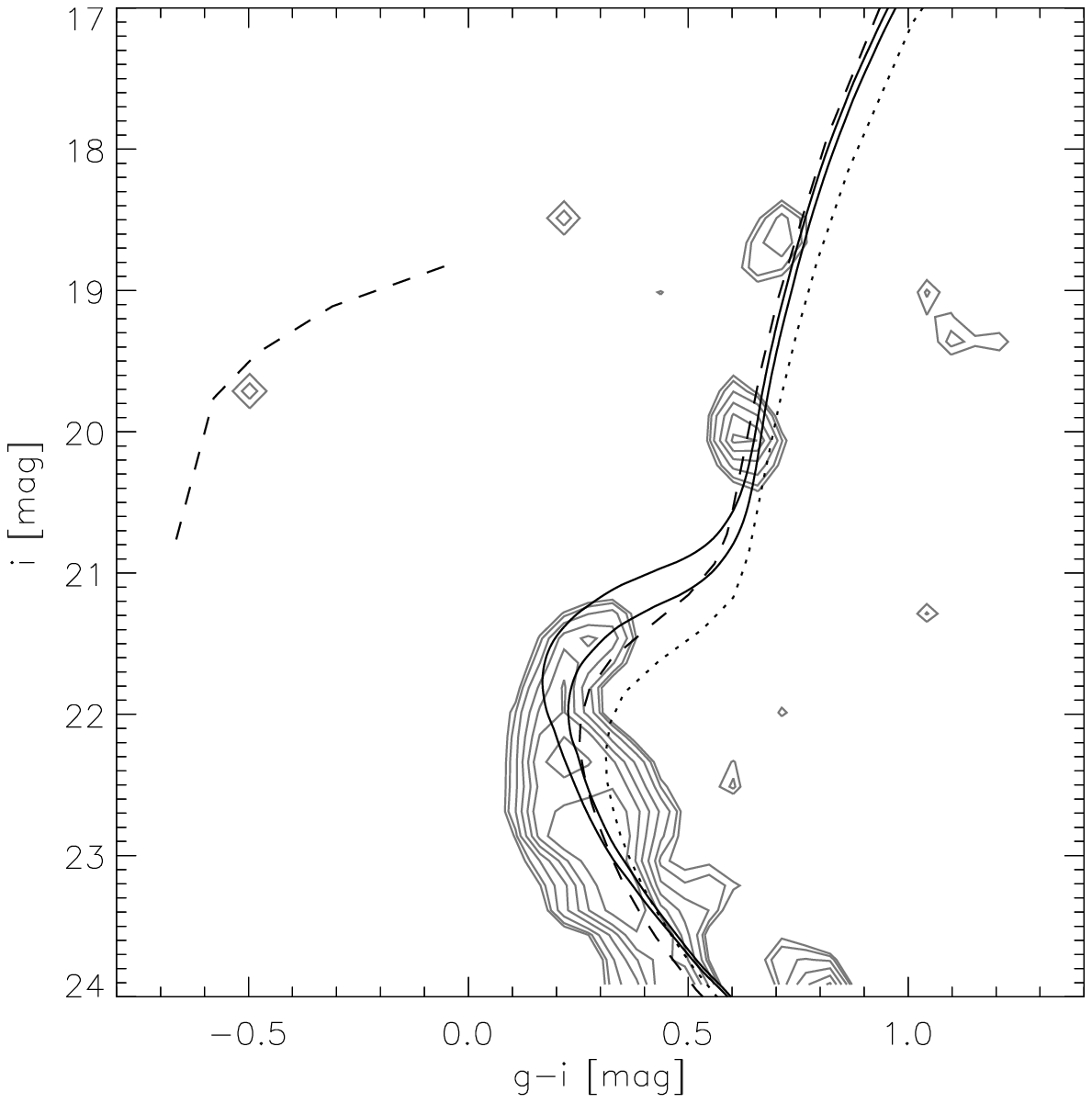}
\end{center}
\caption{CMD for stars within 3$\arcmin$ of the center of Boo~II, with our
spectroscopic target stars indicated (top panel). We distinguish
possible Boo~II members (solid circles) from probable non-members
(open circles). The bottom panel shows a foreground-subtracted Hess
diagram of the same region.
In both panels, the solid black lines show 10 and 12 Gyr Dartmouth
isochrones with [Fe/H]=$-1.8$ dex (Dotter et al. 2008), shifted to a
distance modulus of 18.3.  Also indicated are the fiducials of M92
(dashed; [Fe/H] =$-2.4$ dex) and M13 (dotted; [Fe/H]=$-1.5$ dex) from
Clem et al. (2008). (Note that the HB bottom panel is the fiducial line for 
M92, as opposed to the theoretical isochrones' HB shown in the top panel). 
Our five possible Boo~II members follow the RGB of the isochrone. The
velocity outlier with $v_{\rm HC} = -166 $ km\,s$^{-1}$ (solid square)
falls below the horizontal branches of all the isochrones, and is
therefore likely to be a foreground dwarf. For completeness, the star
with an ambiguous cross-correlation peak is shown as an open square.}
\end{figure}

In order to obtain further insight into the possible association of
each star with the Boo~II overdensity, we compared the most prominent
CMD features with a set of theoretical isochrones.  For this purpose
we performed a by-eye fit of a set of Solar-scaled isochrones from the
Dartmouth group (Dotter et al. 2008) with a metallicity of $-1.8$ dex
(see Sect. 3.3, 3.4) and a reddening of E(B$-$V)=0.03 (Schlegel et
al. 1998), leaving age and distance modulus as free parameters.  We
note that the resulting ``best-fit'' isochrones are not sufficient to
allow us to uniquely characterize the predominant stellar populations
in Boo~II or to derive an accurate distance modulus, but rather enable
us to discuss the plausibility that our targets are Boo~II members.

We find that the  best fit using the isochrone corresponding to
the spectroscopic metallicities ([Fe/H] $=-1.8$ dex) was preferred
over more metal poor and more metal rich tracks, as also indicated by
the globular cluster fiducials in Fig.~6. Neither of these
simultaneously reproduce the MSTO and RGB. In particular, the MSTO of
Boo~II is bluer than the more metal poor GC M92 ([Fe/H]=$-$2.4), which
may also suggest a slightly younger age for Boo~II. Note that the best
age, indicated by the isochrones, is 10--12 Gyr.  In particular, a
good fit is obtained for the RGB, TO and horizontal branch (HB) in the Hess diagram
(bottom panel of Fig.~6), and also of the five individual suggested
member stars (solid circles in Fig.~6, top panel).  Moreover, the good fit 
of the (few) HB stars in the Boo~II CMD assures us that the
distance modulus could be estimated to within 0.2 mag accuracy. The
resulting value of 18.3 mag (46$\pm$4 kpc) is in good agreement with
the measurement of 42$\pm$2 kpc derived by Walsh et al. (2007b). All
in all, our data appear to confirm Boo~II as a moderately old and
metal-poor population (Walsh et al. 2007a,b).

All stars with velocities in excess of $-50$ km\,s$^{-1}$ are well 
separated from the isochrone in Fig.~6, with the exception of  \#13 (with +4 km\,s$^{-1}$). 
The target at $-$165 km\,s$^{-1}$ falls $\sim$1 mag below the HB of any of the 
isochrone models, which indicates that it is most likely a foreground star. 
All of the other five suspected Boo~II member candidates in the radial velocity peak at 
$\sim$$-120$ km\,s$^{-1}$ directly follow the RGB of the ``best-fit'' isochrone, from which we conclude that 
these stars in fact constitute a detection of Boo~II red giant members.

\subsection{Metallicities}
We estimated stellar metallicities from the well-established CaT lines
indicator (e.g., Koch et al. 2006).  To this end, we measured the equivalent
widths (EWs) of each line by fitting a Gaussian plus Lorentz profile
(Cole et al. 2004) and integrating the respective function over the
line band passes defined by Armandroff \& Zinn (1988).  Given the low
S/N ratio of our data, we also chose to integrate the lines
numerically over the standard band passes as a sanity check; both
methods yield consistent values with a mean deviation of 0.01 dex (r.m.s. scatter of 
0.18 dex).  
In practice, we combine the CaT EWs
into a line strength \mbox{$\Sigma W$ = 0.5\,EW(8498) + EW(8542) +
0.6\,EW(8662)}, where we use the calibrations of Rutledge et
al. (1997a; 1997b) onto the Galactic GC scale of Carretta \& Gratton
(1997) to obtain the final metallicities that are stated in Table~1.
That is, \mbox{[Fe/H] = $-$2.66 + 0.42\,[$\Sigma W$ +
0.64\,(V$-$V$_{\rm HB}$)]}.  Here, V is our INT-based V-band magnitude
and V$_{\rm HB}$ denotes the magnitude of the HB.  Given the sparsity
of the HB in the observed CMD (see Figure~1), this value could not be
directly determined from our ob*servations, nor is it well constrained
from the previous photometric studies (Walsh et al. 2007a,b).  Thus we
relied on the locus of the {\em theoretical HB} from the
best-fit isochrone (Sect.~3.2).  One should note that this isochrone
fit relied, in turn, on an adopted metallicity that is {\em a priori}
unknown. As it transpires, however, the best-fit value of V$_{\rm
HB}=18.5\pm0.2$ mag is fairly insensitive to a broader range in
metallicity and the quoted systematic uncertainty reflects this initial ignorance
and also accounts for unknown age and metallicity variations within
the galaxy's stellar populations (Cole et al; 2004; Koch et al. 2006).
Overall, the mean {\em random} error on our metallicities given in Table~1 is 0.16 dex.
Fig.~7 shows the derived metallicities as a function of radial
velocity.
\begin{figure}
\includegraphics[width=1\hsize]{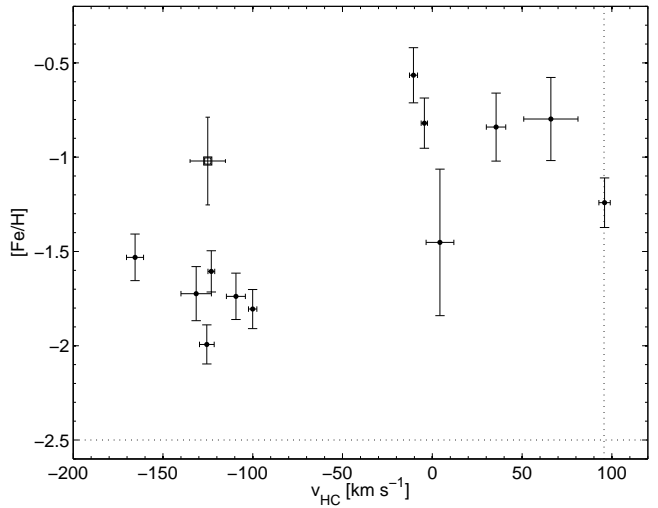}
\caption{Spectroscopic metallicity versus radial velocity for all stars with good velocity measurements. 
The apparent non-member with an ambiguous velocity estimate is
indicated with an open square. The dashed lines delineate the mean
values of velocity and [Fe/H] for Bo\"otes~I (Mu\~noz et al. 2006).}
\end{figure}

Those stars in the velocity clump at $\sim$$-120$ km\,s$^{-1}$ also
show very similar metallicities at around $-1.8$ dex.  On the other
hand, the presumed Galactic foreground contamination at higher
velocities is well separated from Boo~II in that these stars show
systematically higher ``metallicities'' with a broad scatter. Since
the distances to these stars are unknown and the CaT is not a
well-calibrated metallicity indicator for such dwarf stars,
application of the CaT calibrations to the foreground component will
naturally result in an arbitrary metallicity assignment.  The object
that has an ambiguous velocity measurement deviates by $\sim$5$\sigma$
from the mean metallicity, which is consistent with its being a
foreground star.  As an independent test, we also measured the EW of
the gravity sensitive Na doublet lines at 8183, 8195\AA. These are
generally weak in red giants, but show up strongly in dwarf spectra
(e.g., Schiavon et al. 1997) and are thus a powerful dwarf/giant
discriminator (e.g., Koch et al. 2008).  While we do not attempt to
perform a membership separation from this indicator, we note that
those stars with velocities below $-$50 km\,s$^{-1}$ have
systematically lower Na widths (with a mean and 1$\sigma$ scatter of
0.28$\pm$0.36\AA) than those at higher velocities (0.76$\pm$0.34\AA).
With a width of 1.94$\pm$0.51\AA, the object with the spurious
velocity has a Na width which is 2.7$\sigma$ larger than those of the
potential Boo~II giant members, thus strengthening the argument that
it is a foreground dwarf.

\subsection{The characteristics of Bo\"otes~II}

The mean radial velocity and dispersion of {\em all} stars below $-50$
km\,s$^{-1}$ are $-125$ and 19 km\,s$^{-1}$, respectively.  This is an
unrealistically high dispersion and we note that the object with the
most negative velocity in our sample, at $-166$ km\,s$^{-1}$, lies
below the mean velocity value by at least 2$\sigma$, independent of
its inclusion or exclusion from the Boo~II member sample. Considering 
its location in the CMD (Fig.~6), we will treat this star as a
likely non-member for the remainder of this work. The star with the
dubious velocity value does not alter the mean and dispersion by more
than 0.8 km\,s$^{-1}$, but its CMD position and its metallicity argue
against an association with Boo~II and we will not include it in the
further discussions.

This leaves five red giant member candidates with a mean systemic
velocity of ($-$117.0$\pm$5.2) km\,s$^{-1}$ and a velocity dispersion
of (10.5$\pm$7.4) km\,s$^{-1}$, as determined using a maximum
likelihood estimator. If we compare this value to the systemic
velocity for Bo\"otes~I of +95.6$\pm$3.4 km\,s$^{-1}$ (Mu\~noz et
al. 2006), 99.0$\pm$2.1 km\,s$^{-1}$ (Martin et al. 2007), respectively, 
as indicated by the arrow in Fig.~5, any physical connection
between these two systems is firmly ruled out.  Taken at face
value, the dispersion is comparable to the values observed in almost
all of the luminous Local Group dSphs (e.g., Walker et al. 2007;
Gilmore et al. 2007), while the dispersions of the ultrafaint dSphs
(to which Boo~II clearly belongs) are typically 5--7 km\,s$^{-1}$ and
thus systematically lower (Simon \& Geha 2007). The determination of
more velocities of likely Boo~II members is necessary to constrain the
internal kinematic properties of this system.

For our five member stars we find a mean metallicity of
($-1.79\pm0.05$) dex on the scale of Carretta \& Gratton (1997) with a
1$\sigma$-dispersion of 0.14 dex, and 
0.08 dex after correction for measurement uncertainties, respectively.
This spectroscopic mean is higher than the photometric estimate by
Walsh et al. (2007b). By comparison with a fiducial of the old (12
Gyr), metal poor ($-$2.4 dex), $\alpha$-enhanced GC M92, these authors
argue that there is an ``apparent match'' between the cluster and
Boo~II's populations, although a somewhat higher metallicity is not
ruled out by the data, given the sparse RGB.  The most metal poor star
in our sample lies at $-2$ dex and an overlap with a more metal poor
component is thus not excluded.

\section{On the nature of Bo\"otes~II}
The immediate comparison of our derived radial velocities and
metallicties of Boo~II with those of the Bo\"otes~I dSph (dashed lines
in Fig.~6; Mu\~noz et al. 2006; Martin et al. 2007) rules out any physical connection
between these systems. While a difference in mean metallicity of
$\sim$0.7 dex is not critical, it is difficult to envision two
physically associated dwarf galaxies with a relative velocity
difference of $\sim$ 200 km\,s$^{-1}$.

Although it is tempting to identify Boo~II with the classical dSphs
(i.e., an old, metal poor system associated with a single dark matter
halo) based on its stellar population, this interpretation is
complicated by its astrophysical environment, which we now explore.
\subsection{Sagittarius}
At ($\alpha,\delta$)=(209.5$\degr , 12.9\degr$) Boo~II lies in
projection at the edge of the northern, leading arm of the Sagittarius
(Sgr) Stream (Ibata et al. 1997, 2001; Majewski et al. 2003; Belokurov
2006b).  Both the stellar populations (e.g., Newberg et al. 2002;
Majewski et al. 2003; Belokurov et al. 2006b) of the latter feature,
its observed kinematics (Majewski et al. 2004) and subsequent
simulations (Law et al. 2005; Fellhauer et al. 2006) suggest that at
least one wrap of the stream's leading arm passes at a heliocentric
distance of $\sim$40 kpc and that its stars exhibit radial velocities
around $-$100 km\,s$^{-1}$, similar to that found for Boo~II.  This
leads to the natural suggestion that Boo~II may constitute the remnant
of a dSph or star cluster, or a density enhancement (or a tidal dwarf)
that was associated with the Sgr dSph or the progenitor thereof. 

We also note a small excess of stars in the offset control field that coincide 
with location of the main sequence of Boo~II.  A Hess diagram reveals that 
such a feature is fully compatible with noise. If real, that feature would be closer 
than expected for the leading branch of Sgr and might constitute the trailing arm 
that wrapped around the entire Galaxy. There is also photometric evidence for a narrow 
stream that passes the Boo region of the sky and emerges from Sgr 
(e.g., Fig.~1 of Belokurov et al. 2006b).
\begin{figure}
\begin{center}
\includegraphics[width=.9\hsize]{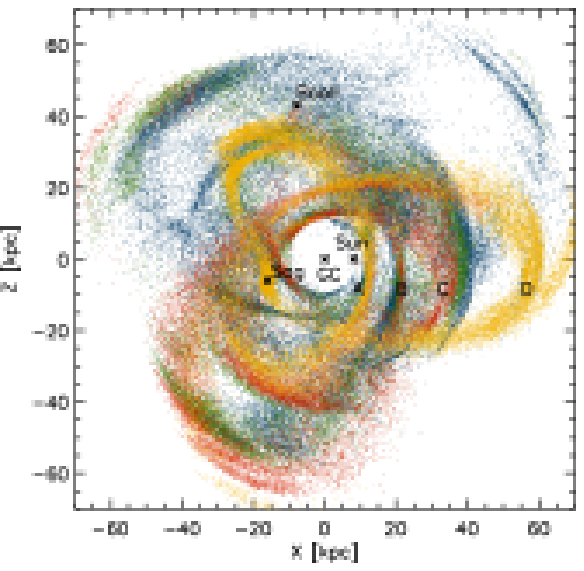}
\includegraphics[width=1\hsize]{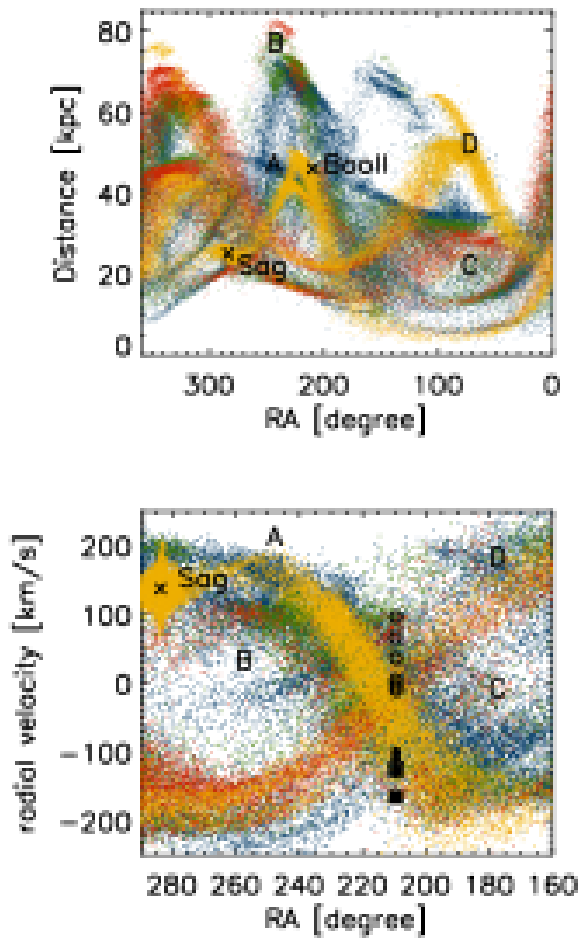}
\end{center}
 \caption{Top and middle panels: Location of Boo~II (black points) on
 the simulations of Fellhauer et al. (2006) of the Sgr Stream. This
 assumes a distance to Boo~II of 46 kpc as estimated in this work. 
 The simulation particles are color coded according to the time they when were lost 
 from Sgr (gold: $<$4 Gyr ago; red: 4--5.7 Gyr; green: 5.7--7.4 Gyr; blue: $>$7.4 Gyr ago). 
 The bottom panel indicates all our good velocity points on the same
 simulations, using the same symbols as in Fig.~6. An association of
 Boo~II with Sgr is thus feasible.}
\end{figure}

In Fig.~8, we overplot the location of Boo~II in Galactic coordinates
(adopting the distance estimate from the present work) and our
measured radial velocities on the simulations of Fellhauer et
al. (2006), which are color-coded by the time when the particles were removed from Sgr. 
\ As this comparison illustrates, both the distance and the
radial velocity range of Boo~II are {\em consistent with an
association with the Sgr Stream}. In particular, its location (top
panel of Fig.~8) suggests a relation with branch ``A'', that is the
``young leading arm'' according to Belokurov et al. (2006b) and
Fellhauer et al. (2006).  The simplistic illustration in the middle
and bottom panels of Fig.~8 appears to favor a stripping of this
feature 4--5.7 Gyr ago (red points) or perhaps 
earlier than 7.4 Gyr ago (blue points in the simulations). 
Given the limited extent of our data
we refrain from overinterpreting any dynamical history of Boo~II.  We
note in passing that the foreground stars at higher velocities also
fall on top of the stream points.

Note that, at an [Fe/H] of $-$1.79 dex, Boo~II is more metal-poor than
the dominant stellar population in the core of the Sgr dwarf, which
has a mean metallicity of $\sim-0.4$ dex (Smecker-Hane \& McWilliam
2002; Bonifacio et al. 2004; Monaco et al. 2005; Sbordone et al. 2007)
but exhibits a broad range in metallicity from $\sim -1.6$ dex to Solar, 
and more metal poor than the predominant field population of the stream (e.g., Monaco
et al. 2007).  On the other hand, based on the significant variations
of the metallicity distributions along the leading arm, Chou et
al. (2007) argue that there must have been a strong metallicity
gradient present in the Sgr progenitor (see also Bellazzini et
al. 2006).  Moreover, Vivas et al. (2005) find that a sample of RR
Lyrae stars in the leading arm is clearly metal poor, at a mean [Fe/H]
of $-1.76$ dex. One possible explanation is thus that Boo~II is a mere
overdensity in the Sgr stream, resembling its known old and metal poor
subpopulation.

Likewise, it is conceivable that Boo~II is a coherent system that has
been tidally stripped from Sgr. There is currently a multitude of GCs
assigned to Sgr and its stream system (Ibata et al. 1997; van den
Bergh 1998; Palma et al. 2002; Bellazzini et al. 2003).  Those GCs
show a broad range in [Fe/H] from the mean of the Sgr dSph ($-0.5$
dex; e.g., Ibata et al. 1997; Brown et al. 1999; Bellazzini et
al. 2002; Cohen 2004; Sbordone et al. 2005) down to $-2$ dex (Da Costa
\& Armandroff 1995) and Boo~II falls in this range.
With a half-light radius of (56$\pm$12) pc, Boo~II is
definitely too extended to be a classical GC and it is also larger
than the confirmed Sgr clusters (with $r_h\la13$ pc).  In the
radius-vs.-$M_V$ plot (e.g., Figure~1 of Gilmore et al. 2007), Boo~II
lies in the gap between the Local Group dSphs and the Galactic
GCs. Although this system is significantly fainter than the GCs, it is
a factor of 5--10 larger in $r_{\rm h}$ and thus comparable in size to
the ultrafaint Com~Ber, Segue~1 (Belokurov et al. 2007) and Willman~1
(Willman et al. 2005b) objects.  Coupled with the large velocity
dispersion found in this work, this argues against Boo~II being a GC,
but rather indicates that it is in fact an ultrafaint, compact
(relative to the more luminous dwarfs) dSph-like object that may have
been stripped from Sgr.
\section{Discussion}
At the time of its discovery, the Bo\"otes~I dSph was the faintest, most
metal poor and most dark matter dominated dSph known (Belokurov et
al. 2006; Mu\~noz et al. 2006; Martin et al. 2007).  Despite its high dark matter content
it is apparently elongated. Fellhauer et al. (2008) argue that this
does not arise due to tidal interactions, but is rather due to
flattening of the progenitor inside an extended dark matter halo.
Adopting a distance to Bo\"otes~I of (66$\pm$3) kpc (Dall'Ora et
al. 2006) and our derived value of 46 kpc for Boo~II, the implied
separation between both these galaxies is (21$\pm$9) kpc and thus
comparable to the present-day LMC--SMC separation (e.g., Gardiner \&
Noguchi 1996).  In contrast to Bo\"otes~I with its high positive
velocity of $\sim 100$ km\,s$^{-1}$ (Mu\~noz et al. 2006; Martin et al. 2007), the fainter Boo~II
dSph exhibits a much lower systemic velocity of $-117$ km\,s$^{-1}$.
A relative velocity difference of $\sim200$ km\,s$^{-1}$ is
inconsistent with a gravitational association between Bo\"otes~I and
II, and is inconsistent with them being on the same orbit through the
Galaxy.

Another important difference between these projected neighbors lies in their 
mean metallicities. With a mean [Fe/H] of $\sim -2.5$,  
Bo\"otes~I represents the most metal poor dSph known to date (Mu\~noz et al. 2006)\footnote{Martin et al. (2007) find a slightly more metal rich mean of [Fe/H]=$-2.1$ dex, which may be related to the adopted 
calibrations of the metallicity scale and extrapolations of these calibrations towards the metal poor 
end.}. 
The fainter Boo~II is found to have a higher mean value of $-1.79$ dex, which  
is compatible with a number of the more luminous dSphs (e.g., Grebel et al. 2003). 

Dwarf galaxies follow well-known metallicity-luminosity and
mass-metallicity relations (Dekel \& Woo 2003; Grebel et al. 2003;
Martin et al. 2007), where the more massive galaxies exhibit higher
metallicity.  This is explicable in terms of the deeper potential
wells of the massive systems, allowing for gas to be retained for a
longer time, leading to more efficient enrichment.  Martin et
al. (2008) estimate a stellar mass in the range of 3.7--7.2$\times
10^4$ M$_{\sun}$ for Boo~II, depending on the adopted Initial Mass
Function.  With the spectroscopic estimate from this work, Boo~II is
significantly more metal rich (by 1 dex) than the value implied by an
extrapolation of the fundamental scaling relation of the more luminous
low-surface brightness galaxies and Local Group dwarfs (Dekel \& Woo
2003). Adding Boo~II to the presently available spectroscopic data of
the ultrafaint dSphs (e.g., Fig.~11 in Simon \& Geha 2007) indicates
that a linear relation between luminosity and metallicity breaks down
for systems fainter than $M_V\ga-5$ mag. In fact, it appears
that such a relation shows an upturn towards higher metallicities for
the faintest systems.

The interpretation of the lowest-luminosity dwarfs in terms of a
tidally stripped remnant (e.g. UMa~II, Com~Ber; Simon \& Geha 2007;
Martin et al. 2008) or association with the Sgr dwarf (Boo~II; this
work) then begs the question of whether these objects constitute a
satellite population distinct from the higher-luminosity dSphs. It is
interesting that both Boo~II and Coma Berenices, the two galaxies
which have half-light radii smaller than the apparent $\sim$$100$ pc
limit identified by Gilmore et al. (2007) and are close to the
Galactic centre, also have mean metallicities which are high relative
to the trend defined by all other dSph galaxies. This is consistent
with these objects being the surviving remnants of parent dSphs which
were originally several magnitudes brighter, and which followed the
luminosity-metallicity trend, and possibly also the minimum-size
relation.

The nature of Boo~II is far from clear.  Given the detection of
distinct, though sparse, CMD features and the kinematic and chemical
evidence derived in this work, we are left with three possible
interpretions of our spectroscopic detection at $-117$ km\,s$^{-1}$:
\begin{enumerate}
\item The overdensity is Boo~II itself and it is an old and 
moderately metal poor dSph. The distinct MSTO in the CMD (Fig.~1;
Walsh et al. 2007a,b) argues in favor of this interpretation.  The
anomalously small half-light radius, which we estimate as 
4.0$\arcmin^{+0.7}_{-0.3}$ (56$^{+8}_{-6}$ pc),  
and anomalously high mean chemical
abundance, together with its Galactic environment, are consistent with
this being the surviving remnant of a larger and more luminous
dSph. Our measured velocity dispersion ($10.5\pm7.4$ km\,s$^{-1}$) is
too uncertain to allow us to derive a reliable mass-to-light ratio, or
even to test the hypothesis that Boo~II is close to internal dynamical
equillibrium.
 \item Comparison with simulations of the Sagittarius stream suggests
 that Boo~II may be a dissolved cluster or a disrupted dSph formerly
 associated with Sagittarius. The metallicities of our Boo~II stars
 are consistent with those found in a broad range of Sgr populations
 and GCs, and is also consistent with the Stream's old and metal poor
 field population (Vivas et al. 2005). We note that the large radius
 of Boo~II is consistent with a disrupting star cluster for only a
 very brief time, so that it is unlikely that such a rare event would
 be observed. Similarly, a star cluster in late disruption has an
 internal velocity dispersion of almost zero km\,s$^{-1}$ (cf. K\"upper et al. 2008). 
 Thus, it is
 unlikely that Boo~II is an unbound, purely stellar system observed at
 a special time.
\item Last, and least comforting, we may have simply not targeted any of the 
real Boo~II stars. 
Given the clumping in velocity, CMD, {\em and} metallicity this
appears unlikely.  Moreover, this option would imply that the galaxy's
real RGB may in fact be even sparser than the apparent feature in the
CMD, yielding an even lower luminosity and mass.
\end{enumerate}
\acknowledgements
AK would like to thank the University of Leicester for hosting a generous stay, during which this 
work was carried out. MIW acknowledges support from a
Royal Society University Research Fellowship.  Based on observations
obtained at the Gemini Observatory, which is operated by the
Association of Universities for Research in Astronomy, Inc., under a
cooperative agreement with the NSF on behalf of the Gemini
partnership: the National Science Foundation (United States), the
Science and Technology Facilities Council (United Kingdom), the
National Research Council (Canada), CONICYT (Chile), the Australian
Research Council (Australia), CNPq (Brazil) and SECYT (Argentina)

\end{document}